\documentstyle[preprint,version2,aps]{revtex}
\begin{document}
\draft
\begin{title}
Order of Two-dimensional Isotropic\\
Dipolar Antiferromagnets
\end{title}
\author{C. PICH and F. SCHWABL}
\begin{instit}
Institut f\"ur Theoretische Physik \\
Physik-Department der Technischen Universit\"at M\"unchen \\
D-8046 Garching, Federal Republic of Germany
\end{instit}
\begin{abstract}
The question of the existence of order in
two-dimensional isotropic dipolar Heisenberg antiferromagnets
is studied. It is shown that the dipolar interaction leads to a
gap in the spin-wave energy and a nonvanishing order parameter. The resulting
finite N\'eel-temperature is calculated for a square lattice by means of
linear spin-wave theory.
\end{abstract}
\narrowtext
The question of order in low-dimensional systems has attracted the interest of
theoretical and experimental physics for a long time. It has been pointed out
already by Bloch \cite{Bloch30} and has been proven exactly \cite{Mermin66}
that long-range order is absent in isotropic two-dimensional Heisenberg
ferromagnets with short-range interaction. The same is true for one- and
two-dimensional He$^4$ \cite{Ferrell65,Hohen67} and antiferromagnets.

In real systems one unavoidably has a dipolar interaction in addition to the
short-range exchange interaction, which breaks the rotational symmetry.
It has been shown by Maleev \cite{Male76} that the $q^2$ dispersion law of the
isotropic ferromagnet is modified such that a finite order parameter exists in
two dimensions. The finite temperature behaviour and in particular the
transition temperature have been calculated by Pokrovsky and Feigelman
\cite{Pok77}.

At first glance one might believe that the dipolar interaction is unimportant
in antiferromagnets due to cancellations because of the alternating
order and it may come as a surprise that this expectation is incorrect.
A first hint that the dipolar interaction can influence antiferromagnetic
behaviour comes from the critical region. There the nonlinear coupling of
fluctuations of the staggered magnetization and the magnetization, which is
no more conserved, leads to a change of the critical dynamic exponent and
of the scaling functions \cite{Fis90}. In the low temperature phase
antiferromagnetic spin-waves involve the coupled precessional motion of
magnetization and staggered magnetization. Since the conservation law for
the magnetization is broken by the dipolar forces also the magnon frequency
becomes finite at wave-vector $\vec q=0$. Thus we will show below that
(i) two-dimensional Heisenberg antiferromagnetic order exists on simple
square lattices due to the dipolar interaction, with the spin orientation
perpendicular to the plane, (ii) the magnon frequency has a gap, the magnitude
of which depends on the dipolar interaction and the exchange interaction,
(iii) the critical field for which the spins rearrange is finite,
(iv) there is a finite N\'eel-temperature which is evaluated.

The Hamiltonian of a dipolar antiferromagnet reads
\begin{equation}
H=-\sum_{l\ne l'}\sum_{\alpha\beta}\left(J_{ll'}\delta_{\alpha\beta}+
A^{\alpha\beta}_{ll'}\right)S_l^{\alpha}S_{l'}^{\beta} -g\mu_B H_0\sum_l
S_l^z
\end{equation}
with spins $\vec S_l$ at lattice sites $\vec x_l$. The first term in brackets
is the exchange interaction $J_{ll'}$ and the second the dipole-dipole
interaction with
\begin{equation}
A^{\alpha\beta}_{ll'}  = -{1\over 2}(g\mu_B)^2\left( {\delta_{\alpha\beta}\over
|\vec x_l -\vec x_{l'}|^3}-{3(\vec x_l-\vec x_{l'})_\alpha(\vec x_l-\vec x_{l'}
)_\beta \over |\vec x_l-\vec x_{l'}|^5}\right)\label{2}.
\end{equation}
Although we are mainly interested in these two terms we have also included a
homogeneous external field $H_0$ along the $z$-axis ($g$ denotes the Land\'e
factor and $\mu_B$ the Bohr magneton). In a two-dimensional system with
additional dipole-dipole interaction the rotational symmetry is broken; thus
the theorem of Hohenberg-Mermin-Wagner \cite{Mermin66,Hohen67} does not apply.

We consider a square lattice in the $xy$ plane with lattice constant $a$ and
the spins orientated alternatingly along the $z$ axis.
By means of the Holstein-Primakoff transformation \cite{Keffer66} the Hamilton
operator can be expressed in terms of the Bose operators ($a_l^{\dag},a_l$)
(neglecting terms higher than bilinear in Eqs.\ \ref{3} and \ref{4})
\begin{equation}
S_l^x  = \sqrt{S\over 2}(a_l+{a_l}^{\dag})~,~~
S_l^y = \mp i\sqrt{S\over 2}(a_l-{a_l}^{\dag})~,~~
S_l^z = \pm(S-{a_l}^{\dag}a_l)\label{3},
\end{equation}
where the upper (lower) sign is for the first (second) sublattice. This
transformation and a Fourier transformation yields
\begin{eqnarray}
H  = & \sum_{\vec q}&\{A_{\vec q}~a_{\vec q}^{\dag}a_{\vec q} + {1\over 2}
B_{\vec q}~(a_{\vec q}~a_{-\vec q}+a_{\vec q}^{\dag}a_{-\vec q}^{\dag})+
\nonumber \\
&& ~~ C_{\vec q}~a_{\vec q}~a_{-{\vec q}-{\vec q}_0}+
C_{\vec q}^{\ast}a_{\vec q}^{\dag}a_{-{\vec q}-{\vec q}_0}^{\dag}+
D_{\vec q}~a_{\vec q}^{\dag}a_{{\vec q}+{\vec q}_0}+D_{\vec q}^{\ast}~
a_{{\vec q}+{\vec q}_0}^{\dag}a_{\vec q}\}\label{4}
\end{eqnarray}
with the coefficients
\begin{mathletters}
\begin{eqnarray}
A_{\vec q}  & = & S(2J_{\vec q_0}-J_{\vec q}-J_{\vec q+\vec q_0})+
S(2A^{zz}_{\vec q_0}-A^{xx}_{\vec q}-A^{yy}_{\vec q+\vec q_0})\\
B_{\vec q}  & = & S(J_{\vec q+\vec q_0}-J_{\vec q}) +
S(A^{yy}_{\vec q+\vec q_0}-A^{xx}_{\vec q})\\
C_{\vec q}  & = &  iSA^{xy}_{\vec q}\\
D_{\vec q} & = & iSA^{xy}_{\vec q}+{1\over 2}g\mu_B H_0.
\end{eqnarray}
\end{mathletters}
In this description (Eq.\ \ref{3}) the primitive cell is the chemical
\cite{Ziman69}, which is half the magnetic. The wave vector
${\vec q}_0={\pi\over a}(1,1)$ represents the antiferromagnetic, staggered
modulation via $e^{i\vec q_0\vec x_l}$.
The $A^{\alpha\beta}_{\vec q}$ are the Fourier transform of the dipole
tensor (Eq.\ \ref{2}) and can be calculated by the method of Ewald
summation \cite{Bonsal77}.

The Hamiltonian (Eq.\ \ref{4}) is diagonalized by a generalized Bogoliubov
transformation with two kinds of creation and annihilation operators
$c_{\vec q}^1,{c_{\vec q}^1}^{\dag},c_{\vec q}^2,{c_{\vec q}^2}^{\dag}$
\begin{equation}
H = E(0) +\sum_{\vec q}\sum_{i=1}^2E_{\vec q}^i~{c_{\vec q}^i}^{\dag}
c_{\vec q}^i
\end{equation}
\begin{equation}
a_{\vec q}=\sum_{i=1}^2u_{\vec q}^i~c_{\vec q}^i+v_{\vec q}^{i^\ast}
c_{-\vec q}^{i{\dag}}+s_{\vec q+\vec q_0}^ic_{\vec q+\vec q_0}^i+
t_{\vec q+\vec q_0}^{i^\ast} c_{-\vec q-\vec q_0}^{i{\dag}}\label{7}
\end{equation}
\[[c_{\vec p}^i~,{c_{\vec q}^{~j}}^{\dag}] = \delta_{\vec p~\vec q}~
\delta_{ij}~,~~[c_{\vec p}^i~,c^{~j}_{\vec q}] = [{c_{\vec p}^i}^{\dag}~,
{c_{\vec q}^{~j}}^{\dag}]=0,
\]
with wave-vectors restricted to the magnetic Brillouin zone. Here $E(0)$ is
the ground-state energy. The spin-wave energies then assume the form
\begin{equation}
{E_{\vec q}^i}^2 = {1\over 2}(\Omega_1 \pm \Omega_2)\label{8}
\end{equation}
with
\[
\Omega_1 =A_{\vec q}^2-B_{\vec q}^2+A_{\vec q+\vec q_0}^2-B_{\vec q+\vec
q_0}^2 +8C_{\vec q}~C_{\vec q+\vec q_0}+2(g\mu_B H_0)^2
\]
and
\[\Omega_2^2 = (A_{\vec q}^2-B_{\vec q}^2-A_{\vec q+\vec q_0}+B_{\vec q+
\vec q_0}^2)^2+
16[C_{\vec q+\vec q_0}(A_{\vec q+\vec q_0}-B_{\vec q+\vec q_0})-C_{\vec q}~
(A_{\vec q}-B_{\vec q})]
\]
\[ \times [C_{\vec q}~(A_{\vec q+\vec q_0}+B_{\vec q+\vec q_0})-
C_{\vec q+\vec q_0}(A_{\vec q}+B_{\vec q})]\]
\[+4(g\mu_B H_0)^2((A_{\vec q}+A_{\vec q+\vec q_0})^2-(B_{\vec q}-B_{\vec q+
\vec q_0})^2).\]
Let us now discuss Eq. (8) in the case of primary interest namely vanishing
external field ($H_0=0$). The dipolar interaction has two effects: first, in
contrast to the isotropic case, the excitation spectrum is no more
degenerate, i.e. two different branches appear. Second it produces an energy
gap for $\vec q\to 0$
\begin{equation}
E_0 = 2S\sqrt{A^{zz}_{\vec q_0}-A^{\rho\rho}_{\vec q_0}}\sqrt{(J_{\vec q_0}
-J_0)-(A^{\rho\rho}_0-A^{zz}_{\vec q_0})}\label{9}
\end{equation}
with
\[A^{\rho\rho}_{\vec k}={1\over 2}(A^{xx}_{\vec k}+A^{yy}_{\vec k}).\]
In Fig. ~\ref{Fig} the dispersion relation is shown for three values
for the ratio of dipolar and exchange energy $\kappa = {(g\mu_B)^2\over
4|J|a^3}$ with isotropic nearest-neighbor exchange interaction ($J<0$). The
two branches can be resolved only for large values of $\kappa$. For more
realistic ratios ($10^{-3}$) the two magnon branches practically coincide,
but a significant deviation from the pure exchange case still remains
in the immediate vicinity of the zone center. The argument of the first
square root in Eq.\ (\ref{9}) for the gap equals the difference of dipolar
energy for out-of- and in-plane staggered orientation and is positive. Thus
stability of the ground state requires
\[J_{\vec q_0}-J_0>A^{\rho\rho}_0-A^{zz}_{\vec q_0}>0.\]
For dipole interaction alone the system would order with the magnetization
in the plane. Thus the dipolar energy difference between in-plane and out-of
plane orientation must be exceeded by the exchange energy in order to
favour the assumed configuration. In particular the gap is proportional to
the square root of the difference of the static energy between the
configurations of in-plane and out-of-plane magnetization. In a
three-dimensional simple cubic lattice the first root in Eq.\ (\ref{9})
vanishes because of the symmetry, but in two-dimensional systems there is a
finite gap for perpendicular antiferromagnetic order. We
note that for sufficiently large exchange energy the gap is the geometric
mean of dipole and exchange energy, which in turn implies that the gap is
much larger than the dipolar energy for $\kappa \ll 1$.

Let us add some comments on the interplay of exchange and dipolar
interaction. The former imposes the antiferromagnetic order while the latter
leads to the orientation perpendicular to the plane and prevents thermal
fluctuations from its destruction. To exhibit more clearly the physical origin
of the energy gap and its principal dependence on dipolar and exchange
interaction we exhibit the equations of motion for the spin components.
Approximating the longitudinal part ($z-$component) by $ S^z_l \approx
Se^{i\vec q_0\vec x_l}$ and specializing to $\vec q =0$, the equations for
the transverse components become
\begin{mathletters}
\begin{equation}
{\dot S}_{0}^x=\quad (A_0-B_0)S_{\vec q_0}^y\label{10}
\end{equation}
\begin{equation}
{\dot S}_{\vec q_0}^{y} = -~(A_0+B_0)S_0^x
\end{equation}
\end{mathletters}
with an analogous set for the $S_0^y$ component. The coefficient on the right
hand side of Eq.\ (\ref{10}) assumes a finite
value in contrast to pure exchange antiferromagnets where $S_{\vec q}^x$ is
conserved. Thus the coupled motion of $S_0^x$ and $S^y_{\vec q_0}$ leads
precisely to the finite energy gap $E_0$ (Eq.\ (\ref{9})) \cite{anmerk}.

{}From the spin-wave energy (Eq.\ (\ref{8})) we can calculate the critical
field
for which the antiferromagnetic N\'eel state gets destabilized by a magnetic
field. It is given by the
field $H_0^c$ for which the energy ($\vec q=0$) vanishes
\begin{equation}
H_0^c = {1\over g\mu_B}E_0.
\end{equation}
Hence the critical field is proportional to the energy gap. In 2D the
anisotropic dipolar interaction stabilizes the antiferromagnetic configuration
in an external field up to the above value.

Now we turn to the evaluation of $T_N$ the transition temperature for
vanishing external field, i.e. the temperature at which the staggered
magnetization vanishes. We use linear spin-wave theory, i.e. interactions
between magnons and temperature renormalization of the magnon energy
are neglected. This approximation is justified at low temperature and should
lead to an order of magnitude estimate of the main dependence on exchange and
dipolar interaction. The staggered magnetization then reads
\begin{equation}
{\rm N}(T)=g\mu_B(NS-\sum_l<a^{\dag}_la_l>)=m_0-{\rm N}_0-{\rm N}_{th}
(T)\label{12}.
\end{equation}
This sum is calculated by means of the transformation (Eq.\ \ref{7}) and
requires the evaluation of the coefficients $u_q,v_q,s_q,t_q$, which are
complicated functions of the the coefficients in Eqs. (5a-5d). The
number of thermally excited magnons $N_{th}(T)$ can be expressed in terms
of the mean number of excitations
\begin{equation}
n_{\vec q}^i=<{c_{\vec q}^i}^{\dag}c_{\vec q}^i>=
\left(e^{E^i_{\vec q}/k_BT}-1\right)^{-1}.
\end{equation}
We now consider isotropic nearest-neighbor exchange ($J<0$) and focuse
only on the limit of small wave-vectors and small dipolar energies
($\kappa \ll 1$). The deviation of the ground-state magnetization due
to thermal excitations then takes the form
\begin{equation}
{\rm N}_{th}(T) = g\mu_B\sum_{i,\vec q}{D\over E_{\vec q}^i}n_{\vec q}^i
\end{equation}
and the deviation originating from the zero-point oscillations
\begin{equation}
{\rm N}_0 =  {1\over 2}g\mu_B\sum_{i,\vec q}\left({D\over E_{\vec q}^i}-
1\right)
\end{equation}
with $D=8S|J|$ and $E_{\vec q}^i\approx\sqrt{D^2a^2q^2/2+E_0^2}$.
The zero point deviation $N_0$  has a finite value \cite{And52} for pure
exchange interaction already, and is effected only neglegibly by the dipole
interaction. The dipole interaction favours the antiferromagnetic order and
leads to a reduction of $N_0$ of order $\kappa$.

Now we turn to $N_{th}(T)$, which is divergent for pure exchange
antiferromagnets implying the absence of antiferromagnetic order in this
case. The small wave-vector approximation in Eqs. (13-15) is not accurate for
temperatures near the phase transition but it is sufficient for our crude
estimate. From Fig. ~\ref{Fig} it becomes clear that the regime of small
wave-vectors is essential for the calculation of the sum.
The existence of a nonvanishing gap makes the sum convergent and allows a
phase-transition (Eq.\ \ref{12}) at a finite temperature $T_N$
\begin{equation}
{\rm N}_{th}(T_N) = m_0 - {\rm N}_0= b'.
\end{equation}
After replacing the sum in Eq. (14) by an integral and the Brillouin zone
by a circle of the same area, the evaluation leads to the following implicit
equation (the upper bound of the integral is set to infinity)
\begin{equation}
e^{E_0\over k_BT_N}-e^{E_0-b\over k_BT_N}=1\label{17}
\end{equation}
for $T_N$ with $b = {\pi D\over 2g\mu_BN}b'$.
In the limit of vanishing gap ($E_0 \to 0$) we recover again the
impossibility of a phase-transition. From Eq.\ (\ref{17}) an asymptotic
solution for small dipole energies can be derived
\begin{equation}
T_N = -{b\over k_B\ln{E_0\over b}} \sim {D\over \ln{D\over E_0}}~~.
\end{equation}

We now compare our results with experiments on $K_2MnF_4$ for which
the spin-wave dispersion  has been measured \cite{Bir73}.
$K_2MnF_4$ is a quasi two-dimensional antiferromagnet with the spin
orientation perpendicular to the ab-plane, a transition temperature
$T_N=42$~K and an exchange energy $|J_1|=8.7$ K ($S=5/2$). An energy gap of
$E_0=7.5$ K is observed. Evaluation of the energy gap via Eq. (9) yields
$E_0=7.6$ K, which is in remarkably agreement with the experimental result.
Solving Eq. (17) with the zero point deviation $N_0 = 0.2Ng\mu_B$ \cite{And52}
our estimate for the transition temperature becomes $T_N= 112$ K which is too
large by a factor of three. It will be lowered if instead of the small wave
vector expansion the correct dispersion relation (Eq. 8) shown in Fig. 1 is
used in the evaluation of $N_{th}$ (Eq.~14). Furthermore  we have
neglected the interaction of magnons, which will be important at higher
temperatures and will lower $T_N$. This could be treated by more elaborate
theories, e.g. \cite{Yab91}, but goes beyond the scope of this paper.

In summary we have shown that two-dimensional antiferromagnetic order is
possible due to the dipolar interaction.

\acknowledgments This work has been supported by the German Federal Ministry
for Research and Technology (BMFT) under the contract number 03-SC2TUM.
We thank E. Frey and W. Gasser for useful comments.

\figure{
The spin-wave dispersion relation (Eq.\ \ref{8}) of pure exchange
antiferromagnets with nearest-neighbour interaction (solid line) and with
additional dipolar interaction ($S=1/2$), for the ratios of dipolar energy
to exchange energy $\kappa={(g\mu_B)^2\over 4|J|a^3}$ along the
${\pi\over a}[\xi,\xi,0]$ direction: $\kappa=0.1$ (dashed), $\kappa=0.01$
(dot-dashed) and $\kappa=0.001$ (dot-dashed-dashed). The splitting of the two
magnon branches is visible only for $\kappa = 0.1$.\label{Fig}}

\end{document}